\documentclass[prb, showpacs, superscriptaddress, twocolumn, floatfix, amsmath]{revtex4}
\usepackage{color, graphicx, amssymb, mathrsfs, multirow}

\date{\today}

\begin{document}
\title{Fluxoid fluctuations in mesoscopic superconducting rings}
\author{Julie A. Bert}  \affiliation{Departments of Physics and Applied Physics, Stanford University, Stanford, CA 94305}
\author{Nicholas C. Koshnick}
\affiliation{Departments of Physics and Applied Physics, Stanford
University, Stanford, CA 94305}
\author{Hendrik Bluhm}
 \altaffiliation[Now at: ]{Department of Physics, Harvard University}
\affiliation{Departments of Physics and Applied Physics, Stanford
University, Stanford, CA 94305}
\author{Kathryn A. Moler} \email{kmoler@stanford.edu}  \ \affiliation{Departments of Physics and Applied Physics, Stanford  University, Stanford, CA 94305}

\begin{abstract}
Rings are a model system for studying phase coherence in one dimension.
Superconducting rings have states with uniform phase windings that
are integer multiples of $2\pi$ called fluxoid states.
When the energy difference between these fluxoid states is
of order the temperature so that phase slips are
energetically accessible, several states contribute to the
ring's magnetic response to a flux threading the ring
in thermal equilibrium and cause a suppression or downturn 
in the ring's magnetic susceptibility as a function of temperature.
We review the theoretical framework for superconducting fluctuations
in rings including
a model developed by Koshnick \cite{koshnick:thesis}
which includes only fluctuations in the ring's phase winding
number called fluxoid fluctuations
and a complete model by von Oppen and Riedel\cite{vonOppen}
that includes all thermal fluctuations
in the Ginzburg-Landau framework.
We show that for sufficiently narrow and dirty rings the two
models predict a similar susceptibility response with a slightly
shifted $T_c$ indicating that
fluxoid fluctuations are dominant.
Finally we present magnetic susceptibility data for rings with different
physical parameters which demonstrate the applicability of our models.
The susceptibility data spans a region in temperature where
the ring transitions from a hysteretic to a non hysteretic response
to a periodic applied magnetic field.
The magnetic susceptibility data, taken where transitions between fluxoid states
are slow compared to the measurement time scale and the ring response was
hysteretic, decreases linearly with
increasing temperature resembling a mean field
response with no fluctuations. At higher temperatures
where fluctuations begin to play a larger role a crossover
occurs and the non-hysteretic data shows a fluxoid fluctuation
induced suppression of diamagnetism
below the mean field response that agrees well with the models.
\end{abstract}


\maketitle

\section{Introduction}

Fluctuations play an important role in the superconducting behavior of
samples of reduced dimensionality:\cite{skocpol:flucts}
they can make electron pairing and long-range phase
coherence occur at different temperatures in unconventional superconductors,
\cite{EmeryPhase} lead to the Berezinskii-Kosterlitz-Thouless transition
\cite{kt_trans} in two dimensions (2D), cause the destruction of long range
phase order in infinitely long one-dimensional (1D) wires, \cite{Rice} and
determine the resistive properties of 1D wires of finite length.
\cite{LangAmb, McCuHalp, giordano_evidence_1988, Lau}

We study the properties of superconducting 1D wires in a model system: uniform
isolated aluminum rings.
An applied magnetic field threading a mesoscopic superconducting ring
drives transitions between states with different phase winding number
(or number of fluxoids).  The process of changing the
fluxoid number by $2\pi$ is called a phase slip.
Changes in fluxoid number also result in
transitions between states with different angular momentum (or vorticity)
with results in jumps in measurable quantities such as the current around
the ring. The nature of superconducting fluctuations in rings
has generated significant interest.
Fluxoid dynamics in individual rings have been probed as
a function of ring
size,\cite{baelus:ringsize, berger:fluxoidjumps, bluhm_multiple_2007} magnetic
field,\cite{berger:fluxoidjumps,pedersen:fluxoidjump,vodolazov:multijump2}
and temperature.\cite{lukens,bluhm_multiple_2007,hernandez:thermaljumps}
The occupation of metastable fluxoid states has
also been measured to determine a crossover
from 1D to 2D behavior in wide rings.\cite{morelle:lpsize,bruyndoncx:lpsize}
Phase slip rates
have been studied in both conventional low $T_c$[\onlinecite{Zhang}] and unconventional high
$T_c$[\onlinecite{kirtley:fluxoiddynamics}] superconducting rings.
Ring inhomogeneities, such as weak links or nonuniform widths, have
been studied as phase slip sites that can impact the ring's current-phase
relationship and fluxoid transitions.
\cite{berger:weaklink, jackel:weaklink, kanda:nonuniform, vodolazov:multijump, vodolazov:nonuniform, vodolazov:multijump2}

Transport measurements have long been used as a probe of superconducting
fluctuations.\cite{Newbower,transport:review}
These experiments measure a voltage
that is directly related to the rate at which phase slips occur.
In contrast, magnetic measurements such as ours are
sensitive to the thermodynamic
equilibrium current in the ring.
Rather than tracking individual fluctuation events,
this paper focuses on the effects
of superconducting fluctuations on the ring's equilibrium supercurrent, $I$, as
a function of applied flux, $\Phi_a$, measured
in a temperature range near the critical temperature, $T_c$.  In
order to reach equilibrium, phase slips
must occur much faster than the measurement time.
We focus on a regime, determined by the ring's physical parameters,
where the distribution and switching between fluxoid states dominate
the ring's equilibrium response to an applied field.
Fluxoid states are the metastable superconducting
solutions which correspond to local minima of the Ginzburg-Landau
free energy functional and can be labeled by an integer phase winding number.
In thermal equilibrium, when the ring's limited thermal energy is
sufficient to cross the barrier between fluxoid states but still small
compared to the saddle point energy, it spends almost no time
near the saddle point solutions that allow phase slips.
As a result, our equilibrium measurement of the ring current depends only on
the energy difference between fluxoid states.

Theoretical work using Ginzburg-Landau (GL)
theory has predicted the current in the presence of
an applied flux threading the ring in the opposite
limit, where fluctuations between states with different
fluxoid numbers are inadequate to
describe the response in the presence of all possible fluctuations
including local phase and amplitude fluctuations.
Ambegaokar and Eckern applied a Gaussian approximation to GL
to predict a mesoscopic persistent current driven by superconducting
fluctuations above $T_c$,\cite{ambegaokar:gaussian1,ambegaokar:gaussian2}
which decreased exponentially with increasing $T$
on the scale of the correlation energy.
However, the Gaussian
approximation, accurate far above $T_c$ where the quadratic term in the
GL free energy dominates,
diverges as $T$ approaches $T_c$.  von Oppen and Riedel
used a transfer matrix approach to GL theory
that accounts for all thermal fluctuations above and below $T_c$
to calculate the supercurrent and correct the divergence
at $T_c$.\cite{vonOppen}
Schwiete and Oreg then proposed a simplification of the
full von Oppen and Riedel (VOR)
theory that makes an analytic prediction for the ring's susceptibility,
$dI/d\Phi_a$, in the limit where the superconducting coherence
length is on the order of the radius,\cite{schwiete} which
provides a simple alternative to solving the VOR model numerically.

Direct measurements of the ring current as a function of applied flux
are useful because it provides access to the
thermodynamic free energy through the derivative
$I=-\partial{F}/\partial\Phi_a$.
While there are also interesting features in the full flux
dependence,\cite{koshnick_fluctuation_2007} the fluctuation response
is well captured by the zero field susceptibility.
Consequently, in this paper we
measure the ring's zero field susceptibility as a function
of temperature, $dI(T)/d\phi|_{\phi=0}$, where $\phi\equiv\Phi_a/\Phi_0$ and
$\Phi_0\equiv{h/2e}$ is the superconducting flux quantum.

A number of different experiments have used susceptibility measurements to
study fluctuations in individual superconducting rings.
\cite{Zhang, bluhm_magnetic_2006, bluhm_multiple_2007,koshnick_fluctuation_2007,jackel:weaklink}  Zhang and Price
studied the phase slip rate and susceptibility as a function of temperature
in a single Al ring. \cite{Zhang}
The ring's geometry and long
mean free path favored amplitude fluctuations that were expected to
support a susceptibility response above
$T_c$. However, the observed susceptibility signal was an order of
magnitude larger than predicted by GL theory.
Koshnick \textit{et al.}\cite{koshnick_fluctuation_2007}
measured the susceptibility of 15 individual rings with long mean free paths
as a function of $\Phi_a$. All rings showed a fluctuation
induced susceptibility response above $T_c$.  The magnitude
of the fluctuation signal was
large in the Little-Parks region, where 
an applied flux suppresses the ring's $T_c(\Phi_a)$ below its zero field value,
$T_c(\Phi_a=0)$. The
susceptibility responses of all the rings in that measurement were well
described by von Oppen and Riedel's complete GL model.\cite{vonOppen}

Dirtier rings or rings with narrower geometries should exhibit
fluxoid fluctuations.
Instead of generating an enhancement in the susceptibility
above $T_c$, fluxoid fluctuations can suppress the superconducting response
well below $T_c$.
This paper focuses on rings that are likely to experience
fluxoid fluctuations. We start by
describing the different thermal fluctuations,
namely phase slips and fluxoid fluctuations,
experienced by our
rings in Sec. \ref{sec_flucts}.  We then
establish the physical conditions that support fluxoid fluctuations in Sec.
\ref{sec_phaseslips}.
We outline a theory\cite{koshnick:thesis}
where a thermal distribution of fluxoid states suppresses the rings
diamagnetism (Sec. \ref{sec_fluxoid}).
Our theoretical analysis concludes by comparing our fluxoid only model to a
complete theory that includes all thermal fluctuations in the GL
framework\cite{vonOppen} (Sec. \ref{sec_VOR}), where we find good
agreement in rings where fluxoid fluctuations dominate the response.
We finally discuss our measurement technique in Sec. \ref{sec_meas}
and present data from two
sets of ring samples with different mean free paths, which we compare to the
theoretical models (Sec. \ref{sec_susc}).

\section{Fluctuation Theory}
\label{sec:fluct_theory}

We begin by describing the different types
of thermal fluctuations encountered
in the GL formalism applied to a ring geometry,
paying particular attention to the difference between phase
slips (processes that change the fluxoid number by moving through a saddle
point in configuration space)
and fluxoid fluctuations (fluctuations in fluxoid number based on the
energy difference between fluxoid states).
We then discuss the onset of thermal equilibrium
by considering the phase slip theory of Langer and
Ambegaokar\cite{LangAmb} as formulated for rings\cite{ZhangThesis, Zhang} in
the context of our actual measurement.
Finally, we introduce fluctuation models that
illuminate how different types of fluctuations
influence the ring's susceptibility response.  These model 
include a mean field
model with no fluctuations,
a model that includes only fluxoid states,
the VOR model including all thermal fluctuations in the GL framework, and
a harmonic oscillator approximation of the VOR model that includes
only quadratic fluctuations.  By comparing these models
we can determine the type of fluctuations that dominate
the response at different
temperatures and for rings
with different physical parameters.

\subsection{Types of Fluctuations}
\label{sec_flucts}

In the context of GL formalism we
introduce a complex order parameter, $\psi(\mathbf{r})$, with an associated
amplitude and phase.  Fluctuations in the amplitude and phase of the order
parameter are deviations in $\psi$ from the mean field
solutions corresponding to local minima of the GL free energy.
Fluctuations play a large role when the thermal energy of the
system allows multiple wavefunctions to
contribute to the ring's properties.

The ring geometry of our samples
imposes a constraint on the order parameter phase.
The periodic boundary
condition requires that the phase be single valued, meaning it must wind by an
integer multiple of $2\pi$ around the ring.
A phase slip is the process of changing the phase winding number
by $2\pi$.
They usually occur by briefly suppressing superconductivity in a
coherence-length-sized section of the ring.\cite{skocpol:flucts}
Rings whose circumference, $L$, is longer than the superconducting
coherence length, $\xi(T)$, have multiple metastable states
that differ by a phase winding or fluxoid number and
their homogeneous superfluid density.  In a
temperature range where phase slips occur within the measurement
time, the ring fluctuates between its minimum energy fluxoid state
and the metastable fluxoid states.\cite{Little}  Such occupation of multiple
fluxoid states or fluxoid fluctuations can contribute to a suppression
of the ring's superconducting response.

\subsection{Phase Slips and Equilibrium}
\label{sec_phaseslips}

The ring's measured response represents its thermal equilibrium response
when phase slips occur at a rate that is fast
compared to experimental time scales ($\sim10$ms),
and no hysteresis is observable.  In this section we derive a
condition for the onset of phase slips and a separate
condition for the onset of fluxoid fluctuations.

We study switching between states driven
by phase slips by adapting the theory of
Langer and Ambegaokar.\cite{LangAmb} Langer and Ambegaokar's treatment
assumes a 1D superconductor where the cross-section is
smaller than the superconductor's coherence length, $\xi(T)$, and
penetration depth,
$\lambda(T)$, so $\psi$ has no radial variation.
This assumption is particularly useful
since exact solutions that minimize the GL free energy exist for 1D systems.
Additionally Langer and Ambegaokar consider a
homogeneous superconductor making phase-slips equally likely at any
point along the ring's circumference.
Langer and Ambegaokar's theory with a correction
to the prefactor by McCumber and Halperin\cite{McCuHalp}
predicts a phase slip rate by calculating the
lowest energy pathway between two fluxoid states as defined by the energy
barrier for the saddle point in wave function configuration space.  Each fluxoid
state, with energy $F_{\rm{min}}$ and phase winding $2\pi{n}$,
represents a stable local minima of the GL free energy functional.
The saddle point energies, $F_{\rm{saddle}}$,
being stationary points of the free energy, must also satisfy the GL equations;
however, these solutions represent unstable configurations.
The phase slip rate, $\Gamma$, then depends exponentially on the energy barrier
${\Delta}F(T,\phi)=F_{\textrm{saddle}}(T,\phi)-F_{\rm{min}}(T,\phi)$.
\begin{equation}
\Gamma=\Omega\exp\left(-\frac{{\Delta}F(T,\phi)}{k_BT}\right)
\label{eq:rate}
\end{equation}
The prefactor, $\Omega = (L/\xi)(\Delta{F}/k_BT)^{1/2}/\tau$, was
corrected by McCumber and Halperin based on arguments using the time dependent
Ginzburg Landau model,\cite{McCuHalp} where $\tau = \pi h / 8 k_B(T_c -T)$
is the relaxation time. For phase slips to occur at a rate of
$100\,$Hz, a rate that is approximately equal to our measurement time, the
energy difference $\Delta{F}/k_BT$ may not exceed $\sim{20}$.
Although the prefactor
is large, $\Omega\approx5\times{10}^{11}-5\times{10}^{12}\,$Hz,
the exponential
dependence overwhelms changes in the prefactor which can therefore be ignored.
We find an approximate condition for the onset of phase slips
from calculations of the energy barrier $\Delta{F}$.

We find expressions for $F_{\rm{min}}$ and $F_{\rm{saddle}}$
which represent stationary points of the GL free energy.
One dimensional Ginzburg-Landau theory introduces the GL free energy
functional in the presence of a magnetic field represented by the
vector potential $\vec{A}$ as

\begin{eqnarray}
\lefteqn{F[\psi(x)]=\int\left[\alpha|\psi(x)|^2+\frac{1}{2}\beta|\psi(x)|^4
\right.} \nonumber\\
 & & +\left.\frac{\hbar}{2m^*}\left|\left(\vec{\nabla}-
\frac{ie^*\vec{A}}{\hbar}\right)\psi(x)\right|^2\right]d^3x.
\label{eq:GLfree}
\end{eqnarray}
$\alpha$ and $\beta$ both depend on $T$, and
$\alpha^2/\beta = B_c(T)^2/\mu_0$ is related to the superconducting critical
field, $B_c(T)$. $e^{*}$ and $m^{*}$ are the charge and mass of the Cooper
pairs and $\mu_0$ is the permeability of free space.

We look for stable solutions that locally minimize the Ginzburg Landau
free energy functional. In a homogeneous one dimensional ring
fluxoid states have free energies
\begin{equation}
F_{\rm{min}}(T,\phi) = -F_c(T) \left( 1 - \frac{\xi(T)^2}{R^2} (\phi - n)^2 \right)^2,
\label{eq:F_ff_gl}
\end{equation}
where the critical field, and the ring volume ($V=2\pi{R}wd$)
determine the ring's total condensation energy ($F_c(T)=V B_c(T)^2/2\mu_0$).
$w$ is the ring width and $d$ is the thickness.
The dependence on  $\xi(T)/R$ accounts for the
suppression of the superfluid density by the phase gradient
around the ring with coherence length $\xi(T)$ and radius $R$.
The Aharonov-Bohm flux,
$\phi = \Phi_a/\Phi_0$, can be transformed into a
shift in the boundary conditions for a wave function in a ring,
\cite{byers_theoretical_1961} and therefore contributes to the energy in
the same way as $n$.

To find the saddle point energies,
Zhang applied Langer and Ambegaokar's phase slip theory to a ring
geometry,\cite{ZhangThesis} and by used the approximation $L\gg\xi(T)$ to
calculate the saddle point energy
\begin{eqnarray}
\lefteqn{F_{\textrm{saddle}}(T,\phi) =}\nonumber\\
&&F_c(T) \left(\frac{8\sqrt{2 \delta(T,\phi,n)}}{3} \frac{\xi(T)}{L} - \frac{ (2+\delta(T,\phi,n))^2}{9} \right),
\label{eq:saddle}
\end{eqnarray}
where
$L=2\pi{R}$ is the ring's circumference and
$\delta(T,\phi,n)$ is the normalized difference between the square of the order
parameter amplitudes near and far from a phase slip event.
$\delta(T,\phi,n)$ is a real number between 0 and 1 that satisfies the relation
\begin{equation}
2\pi n = \sqrt{\frac{1-\delta}{3}}\frac{L}{\xi(T)} + 2\tan^{-1}\left(\sqrt{\frac{3\delta} {2(1-\delta)}}\right) + 2\pi\phi.
\end{equation}
For $\phi=n+1/2$, $\delta = 1$.

We are interested in a regime where $L \gg \xi(T)$,
and $\delta$ remains close to one for moderate $n$.
Using the substitution $\kappa=\sqrt{1-\delta}$ and expanding to
lowest order in $\kappa$, we arrive at a simplified expression for $\delta$.
\begin{equation}
\delta(T,\phi,n) = 1 - \left( \frac{\sqrt{3}\pi(2n-2\phi-1)}{\frac{L}{\xi(T)}-2\sqrt{2}}\right)^2
\end{equation}
\begin{figure}[b]
\centering
\includegraphics[scale=1]{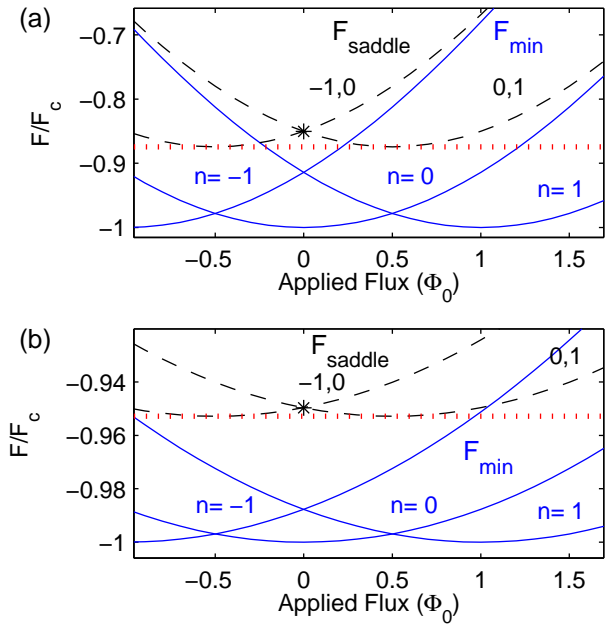}
\caption[Fluxoid and saddle points energies versus applied flux.]{
(Color Online) 
Free energy of fluxoid states (blue solid line, Eq. \ref{eq:F_ff_gl})
and saddle point energies (black dashed line, Eq. \ref{eq:saddle}) as a function
of applied flux, in units of the condensation energy, $F_c$.
The red dotted line is the minimum value of $F_{\rm{saddle}}$
which is approximately equal to $3.8$ times the condensation energy 
for a ring section of length $\xi$.
({\bf a})  The respective energies when $L / \xi(T) = 30$.
({\bf b})  The respective energies when $L / \xi(T) = 80$.
The energy barrier between adjacent fluxoid states,
$\Delta{F^{\pm1,0}_{\rm{min}}}= F^{n=\pm1}_{\rm{min}} - F^{n = 0}_{\rm{min}}$,
scales differently than the saddle point energy $\Delta{F}$.
\label{fig:EnIvsFlux}}
\end{figure}

Returning to the energy expression, Eq. (\ref{eq:saddle}), we find
minima of $F_{\textrm{saddle}}$ at $n-\phi=1/2$
\begin{equation}
{\rm{min}}(F_{\textrm{saddle}})= F_c(T)\left( \frac{8\sqrt{2}}{3}\frac{ \xi(T)}{L} - 1 \right).
\label{eq:minFs}
\end{equation}

The energy expressions for $F_{\rm{min}}$ (Eq. (\ref{eq:F_ff_gl}))
and $F_{\rm{saddle}}$ (Eq. (\ref{eq:saddle})) are
plotted as a function of flux for different fluxoid number $n$ in
Fig. \ref{fig:EnIvsFlux}.  The minimum value of $F_{\rm{saddle}}$ given
in Eq. (\ref{eq:minFs}) is also plotted.
The plots demonstrate how an increase in the ratio $L/\xi(T)$
decreases the variation of both $F_{\rm{min}}$ and $F_{\rm{saddle}}$
with applied flux.  For $L/\xi(T)=30$ the variation in
$\Delta{F}(\phi)$ between $\phi=0$ and $\phi=1/2$ is $30\%$ of
$\Delta{F}(\phi=0)$.  This variation decreases to $12\%$ for
$L/\xi(T)=80$.

As $L/\xi(T)$ becomes increasingly large, in rings with
long circumferences,
the flux dependence of both $F_{\textrm{saddle}}(T,\phi)$ and
$F_{\rm{min}}(T,\phi)$ becomes less
important to the energy barrier
$\Delta{F}(T,\phi)$.
This decreasing dependence on flux justifies the approximation
that neither $F_{\rm{min}}$ nor $F_{\rm{saddle}}$ depends on the flux
in rings where $L\gg\xi(T)$. Approximating
$F_{\textrm{saddle}}(T,\phi)\approx {\rm{min}}(F_{\textrm{saddle}})$
and $F_{\rm{min}}(T,\phi)\approx {\rm{min}}(F_{\rm{min}})$, leaves
$\Delta{F}(T) \approx 3.8 wd\xi B_c(T)^2/2\mu_0$, which
is also independent
of flux. The exponential dependence of the phase slip rate combined
with the large attempt frequency $\Omega$ in
Eq. (\ref{eq:rate}) imply phase slips occur when
$\Delta{F}(T)\lesssim 20k_BT$. Qualitatively this result
makes sense because it states that phase slips occur when there is
sufficient thermal energy to suppress the order parameter on the
scale of the coherence length.
This energy scale differs from the ring's total condensation energy, $F_c$,
which is proportional to the volume of the entire ring.

In contrast to the phase slip energy we just derived,
the we obtain the energy associated with each fluxoid state by
expanding the mean field GL free energy expression, Eq. (\ref{eq:F_ff_gl}),
to lowest order in $\xi(T)/R$.
\begin{equation}
F^{n}_{\rm{min}}(T,\phi) = \frac{I^0(T)\Phi_0}{2} (\phi-n)^2  - V {\frac{B_c(T)^2}{\mu_0}}
\label{eq:energy}
\end{equation}
where
\begin{equation}
I^0(T)=\frac{2V{B_c(T)}^2\xi(T)^2}{\Phi_0\mu_0{R^2}}=\frac{4\pi^2k_BTL^2}{\Phi_0\xi(T)^2\gamma(T)}
\label{eq:Izero}
\end{equation}
We introduce the parameter $\gamma(T)$ which describes the inverse phase
stiffness of the ring.  Rings with a larger $\gamma(T)$ are more
likely to exhibit fluxoid fluctuations.  $\gamma(T)$ is discussed in more
detail in section \ref{sec_VOR}.  

When the energy difference between the lowest energy fluxoid states,
$\Delta{F^{\pm1,0}_{\rm{min}}}=F^{n = \pm1}_{\rm{min}}-F^{n=0}_{\rm{min}}$,
is comparable to the temperature, multiple fluxoid states contribute
to the ring's equilibrium response.

The criterion for phase slips
($\Delta{F}(T)\approx 3.8 wd\xi B_c(T)^2/2\mu_0$)
scales differently with ring radius compared to
the energy difference between fluxoid states
($\Delta{F^{\pm1,0}_{\rm{min}}}(T)\approx 4\pi{wd\xi}B_c(T)^2/2\mu_0(\xi/R)$)
that determines onset temperature for fluxoid fluctuations.  
As stated above phase slips begin when
$\Delta{F}(T)\lesssim20k_BT$, while fluxoid fluctuations
onset for $\Delta{F^{\pm1,0}_{\rm{min}}}\lesssim6k_BT$.
See section \ref{sec_fluxoid} Eq. (\ref{eq:ff_limit}).  
These two types of fluctuations onset at the
same temperature when $\xi(T)/L=0.014$.  Therefore,
in the limit where $L \gg \xi(T)$
(specifically when $\xi(T)/L<0.014$), the
condition for fluxoid fluctuations
is already satisfied when phase slips begin to
occur. By applying the fluxoid condition, Eq. (\ref{eq:ff_limit}),
and making an approximation that $T\approx{T_c}$ we can remove the 
temperature dependence and rewrite the
condition as $\gamma(T_c) > 16,000$.
As a result, in the longest wires with the largest $\gamma(T)$
a suppression of the response due to
fluxoid fluctuations is expected as soon as the ring is able to reach
thermal equilibrium.  

In the opposite limit, for small clean rings, if the condition
$\Delta{F^{\pm1,0}_{\rm{min}}}\lesssim 6k_BT$ is not met by the time
$\Delta{F}(T)\approx{k_BT}$ then thermal occupation of the saddle point
solutions and other higher energy states become significant.
This occurs in rings where $\gamma(T_c)<40$.
When fluxoid fluctuations onset they will not dominate the 
fluctuation response and their effect on the zero-field
susceptibility is negligible.

Using Eq. \ref{eq:rate} we can predict the onset of phase slips in
aluminum rings.  A ring with a mean free path $\ell_e = 4\,$nm has
$\lambda(0) \approx 800\,$nm and 
$\xi(0) \approx 85\,$nm.\cite{bluhm_magnetic_2006}   Using standard
temperature dependencies for these parameters,$\xi(T)\propto(1-t)^{-1/2}$ and
$\lambda(T)\propto(1-t)^{-1/2}$ where $t=T_c/T$,\cite{tinkham_SC} a ring of this
material with $w = 80\,$nm and $d = 40\,$nm will have switching occur on
experimental time scales down to 1.1 K when $T_c \approx 1.24\,$K,
while a ring
with $d = 1\,$nm will continue to have phase slips down to 0.45 K.

We have shown that the onset temperature for phase slips,
$\Delta{F}\propto\xi(T)$,
and for fluxoid fluctuations, $\Delta{F^{\pm1,0}_{\rm{min}}}\propto\xi(T)^2/R$,
scale differently with the ring radius.
However, the presence of phase slips is a precondition
for fluxoid fluctuations.
The phase slip rate, Eq. (\ref{eq:rate}), depends on
the temperature and the energy difference
between the saddle point energy and the minimized GL energy.
In the $L\gg\xi(T)$ limit, the rate is independent of applied flux and
the ring circumference, $\Gamma\propto\exp(-3.8wd\xi(T)B_c(T)^2/2\mu_0k_B{T})$.
This phase slip rate
and the measurement time sets our criterion
for equilibrium.
Namely a ring's measured response represents its
equilibrium response when phase slips occur
faster than the time over which the flux is swept through the transition
region.

In the following sections we explore the effects of
fluctuations on the ring's response in thermal equilibrium.
Each of the models presented below includes a
different set of fluctuations.  By comparing the model predictions
for different ring parameters we can pinpoint the effect
of different fluctuations on ring response and set a physical regime where
each of the fluctuations will dominate.  Specifically
we find that in rings with weak phase stiffness a model including only
fluxoid fluctuations accurately reproduces the ring response except for
an apparent shift in $T_c$ indicating
that fluxoid fluctuations dominate.

\subsection{Fluxoid Number Distribution Model}
\label{sec_fluxoid}
We start with a model that includes only fluxoid fluctuations:
fluctuations caused by integer winding or unwinding
of the phase of the order parameter.  This
model is not complete because it does not include nonuniform
phase fluctuations or
amplitude fluctuations.  Put another way this model includes
only the large fluctuations
between local minima in the GL free energy
(see Fig. \ref{fig:EnIvsFlux}), and
ignores all the small fluctuations about each local minimum
as well as the saddle points and intermediate states.
It is instructive to
develop this model because comparisons between this fluxoid
only model and more complete
models shed light on what portion of the fluctuation
response of a ring is due to solely to fluxoid fluctuations.
As discussed in the previous section, we expect this model
to be valid for rings with $\gamma(T_c)>40$.

We return to the mean field 1D GL free energy,
Eq. (\ref{eq:F_ff_gl}),
which is related to the
ring current by $I=-\partial{F}/\partial\Phi_a$.
Taking the derivative yields an expression for
the ring current of a state with $n$ fluxoids.
\begin{equation}
I_n(T,\phi)=I^0(T)(\phi-n)\left(1-\frac{\xi(T)^2}{R^2}(\phi-n)^2\right)
\label{eq:GL_current}
\end{equation}
where $I^0(T)$ was given in Eq. (\ref{eq:Izero}).
$I^0(T)(\phi-n)$ is the Meissner response which decreases linearly
with increasing temperature close to $T_c$, while the cubic term
arises from pair-breaking.

The energy associated with each fluxoid current state, $F^{n}_{\rm{min}}$,
was derived in the previous section, Eq. (\ref{eq:energy}).
If phase slips occur at a high
enough rate, so that the metastable fluxoid states
are in thermal equilibrium as discussed in the previous section,
we can model\cite{bluhm_magnetic_2006,skocpol:flucts}
the resulting current response as a Boltzmann distribution of fluxoid states.
\begin{equation}
I_{F}(T,\phi) = \frac{\sum_n{I_n}(T,\phi)\exp{(-F^{n}_{\rm{min}}(T,\phi)/k_BT)}}{\sum_n\exp{(-F^{n}_{\rm{min}}(T,\phi)/k_B T)}}.
\label{eqn:fluxfluct}
\end{equation}
We call the total ring current generated by fluxoid states $I_F$ to
distinguish it from the total ring current including all fluctuation
states that will be presented in the next section.
The susceptibility response of the ring at zero applied flux,
$dI(T)/d\phi|_{\phi=0}$ gives us access to the
thermodynamic free energy.
In our rings $\xi(T)/L\ll{1}$, so we make
the approximation $I_n(T,\phi){\approx}I^{0}(T)(\phi-n)$.
The derivative of the total ring current at $\phi = 0$ is
\begin{equation}
\frac{dI_{F}(T)}{d\phi}\bigg|_{\phi = 0} = I^0(T) \left(1 - \frac{\sum_n2\sigma{n^2}\exp(-\sigma n^2)}{\sum_{n}\exp(-\sigma{n^2})}\right)
\label{eqn:suscfluxfluct}
\end{equation}
where $\sigma \equiv I^0(T) \Phi_0 / 2 k_BT$.
Eq. (\ref{eqn:suscfluxfluct}) shows that
including a distribution of fluxoid states reduces the
ring's susceptibility response from the mean field value, $I^{0}(T)$.
However, the ring's
superfluid density is not reduced. The second
term in Eq. (\ref{eqn:suscfluxfluct}) is proportional to the
RMS fluctuation of the fluxoid number, $n$.  The magnitude of the
reduction in susceptibility depends on $\sigma$.
When $\sigma$ is large, terms with
$n \ne 0$ are small and
the susceptibility is approximately equal to the mean field value.
When $\sigma$ is small, the $n = \pm1$ terms begin to play a significant role.
We define a criterion\cite{koshnick:thesis} for fluxoid fluctuations
to reduce the Meissner response by more than 5\% when
\begin{equation}
\frac{dI_{F}(T)}{d\phi} \approx I^0(T) < \frac{12k_BT}{\Phi_0},
\label{eq:ff_limit}
\end{equation}
as long as phase slips occur at a sufficiently high rate.
We used this criterion in section \ref{sec_phaseslips}
to compare the onset of fluxoid fluctuations and phase slips.
In plots of the susceptibility vs. temperature we observed
a suppression below the mean field value for susceptibilities below
this cutoff.  This downturn in the susceptibility signal, which
occurs at $T$ less than $T_c$, is a hallmark of the
suppression of the diamagnetic response by fluxoid fluctuations.
More specifically, as stated in section \ref{sec_phaseslips}
we expect a visible downturn in rings with $\gamma(T_c)$ values between
$40$ and $16000$.  Rings with $\gamma(T_c)<40$ will not show a downturn
as higher energy fluctuations overwhelm fluxoid fluctuations.
In rings with $\gamma(T_c)>16,000$ fluxiod fluctuations will occur
as soon as thermal equilibrium is reached. These rings will exhibit
a response that is already suppressed below the mean field value for
all data taken in thermal equilibrium.
The next sections will be devoted
to developing more complex models which include a more complete
set of thermal fluctuations.

Thus far, we have considered a fluxoid model that predicts
the existence of the downturn in
susceptibility below $T_c$.
In some rings, near $T = T_c$, the $R \gg \xi(T)$ assumption we made
to obtain Eq. (\ref{eq:energy}) and (\ref{eqn:suscfluxfluct})
is not strictly valid.   As a result,
the energy between successive metastable states can no longer be
approximated by the expansion in Eq. (\ref{eq:energy}).  By including
the quartic term of Eq. (\ref{eq:F_ff_gl}), the GL free energy
vanishes rather than increasing indefinitely for $\phi - n > L/\xi(T)$.
Thus, the Boltzmann distribution, Eq. (\ref{eqn:fluxfluct}), is also
not well defined because summing over all $n$ leads to a divergent
denominator. The numerator on the other hand remains finite since
states with $\phi - n > L/\xi(T)$ do not contribute.
Furthermore, our treatment thus far has ignored phase fluctuations that
are not uniform around the ring and all amplitude fluctuations.

\subsection{von Oppen and Riedel Model}
\label{sec_VOR}
To address these issues, we compare our simple fluxoid model to the model
of von Oppen and Riedel,\cite{vonOppen} VOR, which generates numeric
solutions that include all thermal fluctuations within the
GL framework in homogeneous rings.
Applying a harmonic oscillator approximation to the VOR model, discussed
in the next section, provides a
direct mathematical connection between the VOR and
fluxoid model discussed in the previous section.

Following von Oppen and Riedel,\cite{vonOppen} we begin with
the expression for the GL energy functional given in
Sec. \ref{sec:fluct_theory} Eq. (\ref{eq:GLfree}).
We map the free energy onto a one dimensional ring geometry
with no lateral variation of the order parameter so
$\psi(r,z,\theta)=\psi(\theta)$ and $dx^3=wdRd\theta$.  We
can then redefine
$Rd\theta$ as $dx$.

We rewrite Eq. (\ref{eq:GLfree}) using reduced variables
$\psi(x)=\bar{\psi}(\bar{x})\sqrt{|\alpha|/\beta}$, $\bar{\nabla}=\xi\nabla$,
and $x=\bar{x}\xi$.  $\xi(T)$ is the superconducting coherence length and is
given by $\xi(T)=\hbar/\sqrt{2m^{*}\alpha}$.

\begin{eqnarray}
\lefteqn{F[\bar{\psi}(\bar{x})]=} \nonumber\\
&&E_0(T)k_BT\int^{\Lambda(T)/2}_{-\Lambda(T)/2}
\left[\eta|\bar{\psi}(\bar{x})|^2 + \frac{1}{2}|\bar{\psi}(\bar{x})|^4 \right . \nonumber\\
&&+\left.\left|\left(\bar{\nabla}-\frac{2\pi{i}}{\Lambda(T)}\phi\right)
\bar{\psi}(\bar{x})\right|^2 \right] d\bar{x}
\label{eq:VOGL}
\end{eqnarray}
$\eta$ is $+1(-1)$ for temperatures above (below) the superconducting
critical temperature $T_c$.  $\Lambda(T)$ is the reduced circumference
$\Lambda(T)=L/\xi(T)=\sqrt{8{\pi}{k_B}|T-T_c|/E_c}$ and
$E_0(T)k_BT=wd{\xi(T)}B_c(T)^2/\mu_0$
is the condensation
energy of a ring section of length $\xi(T)$.
The correlation energy
for the ring, $E_c=\pi^2\hbar{v_f}\ell_e/3L^2$, includes the
mean free path, $\ell_e$, and
the fermi velocity, $v_f$,
which is $2.03\times10^{6}\,$m$/$s in aluminum.
$E_0(T)$ can also be written as
\begin{equation}
E_0(T)=\frac{(2\pi)^{5/2}}{21\zeta(3)}\left(\frac{k_B|T-T_c|}{E_c}\right)^{3/2}
\frac{E_cM_{\rm{eff}}}{k_BT},
\label{eq:E0}
\end{equation}
where $\zeta(3)=1.021$ is the Riemann zeta function.
$M=k_f^2wd/4\pi$ is the number of transverse channels.
$k_f$ is the fermi wave vector, which for an aluminum ring is
$k_f = 1.75\times10^{10}\,$m$^{-1}$.
Including disorder results in an effective number of
channels, $M_{\rm{eff}}=M\ell_e/L$.

We obtain the thermodynamic expression of the current from the
flux derivative of the ring's partition function.
\begin{equation}
I(T,\phi) = -k_BT{\frac{1}{Z_{sc}}}{\frac{\partial}{\partial\Phi_a}}Z_{sc}
\label{eq:thermocurrent}
\end{equation}
The partition function is the path integral of the GL energy.
\begin{equation}
Z_{sc} = \int[d\bar{\psi}(\bar{x})][d\bar{\psi}^{*}(\bar{x})]\exp\left(\frac{-F[\bar{\psi}(\bar{x})]}{k_BT}\right)
\label{eq:partition}
\end{equation}
The VOR model uses a transfer matrix technique \cite{scalapino}
to map the Ginzburg-Landau path integral partition function, Eq.
(\ref{eq:partition}), onto another partition function
\begin{equation}
Z = \sum_{l = -\infty}^{\infty} \exp(-i 2 \pi l \phi) \sum_{n=0}^{\infty} \exp(- 2E_0(T)\Lambda(T) \mathscr{E}_{n,l} )
\label{eqn:voppZff}
\end{equation}
where $\mathscr{E}_{n,l}$ are the eigenvalues of the fictitious 2D single-particle Hamiltonian,
\begin{equation}
H = -\frac{1}{8E_0(T)^2} \nabla^2 + \frac{1}{2}\eta{\vec r}^2 + \frac{1}{4}{\vec r}^4.
\label{eqn:voppHff}
\end{equation}
We define $\vec{\rho} = (2E_0(T))^{1/3}\vec{r}$ and rewrite Eqs.
(\ref{eqn:voppZff}) and (\ref{eqn:voppHff}) to emphasize the parameter
$\gamma(T)$.\cite{koshnick_fluctuation_2007}
\begin{eqnarray}
&&Z = \sum_{l = -\infty}^{\infty} \exp(-i 2 \pi l \phi) \sum_{n=0}^{\infty} \exp(- \gamma(T)^{1/3}\mathscr{E}_{n,l} )
\label{eq:nVORpart}\\
&&H = -\frac{1}{2} \nabla^2 + \frac{1}{2}\frac{\Lambda(T)^2}{\gamma(T)^{2/3}}{\vec \rho}^2 + \frac{1}{4}{\vec \rho}^4
\label{eq:nVORhamiltonian}
\end{eqnarray}
The temperature dependence is set by the coherence
length through the relation $\Lambda(T) = L/\xi(T)$.
The parameter
\begin{equation}
\gamma(T) \equiv \frac{\Lambda(T)^3}{2E_0(T)} = \frac{42 \zeta(3)}{ \pi}\frac{  k_B T}{ M_{\mathrm{eff}} E_c}
\label{eq:gamma}
\end{equation}
can be thought of as a measure of
inverse phase stiffness, which determines the type of fluctuations
that dominate the ring's susceptibility
response.\cite{koshnick_fluctuation_2007}
The definition of $\gamma$ introduced in Koshnick
\textit{et al.}\cite{koshnick_fluctuation_2007} made the approximation
$T\approx{T_c}$.  The larger temperature range explored in this paper
makes it necessary to reintroduce the $T$ dependence.
We use the relation for $I^0(T)$ given in the second 
part of Eq. (\ref{eq:Izero}) to compare
the VOR model to the mean field and fluxoid models.

Eqs. (\ref{eq:nVORpart}) and (\ref{eq:nVORhamiltonian}) can be solved
numerically.  The Hamiltonian can be rewritten as a harmonic oscillator
with a quartic perturbation.  We can then write matrix elements
in terms of the coefficients and diagonalized numerically
to find the eigenvalues.\cite{ZhangThesis,Bell}  The eigenvalues
are used in the partition function, Eq. (\ref{eq:nVORpart}), and
substituted into the thermodynamic equation for the current,
Eq. (\ref{eq:thermocurrent}),
to generate the full current response.  We find the zero-field
susceptibility by taking a derivative with respect to applied flux at $\phi=0$.

Analytic solutions can be instructive, and
as a result it is useful to find approximations
to the full VOR model that are valid over some set of ring parameters
or temperatures.  One such approximation is to ignore the quartic
perturbation to the Hamiltonian, which then takes the form of a
simple harmonic oscillator.

\begin{figure}[t]
\centering
\includegraphics[scale=1]{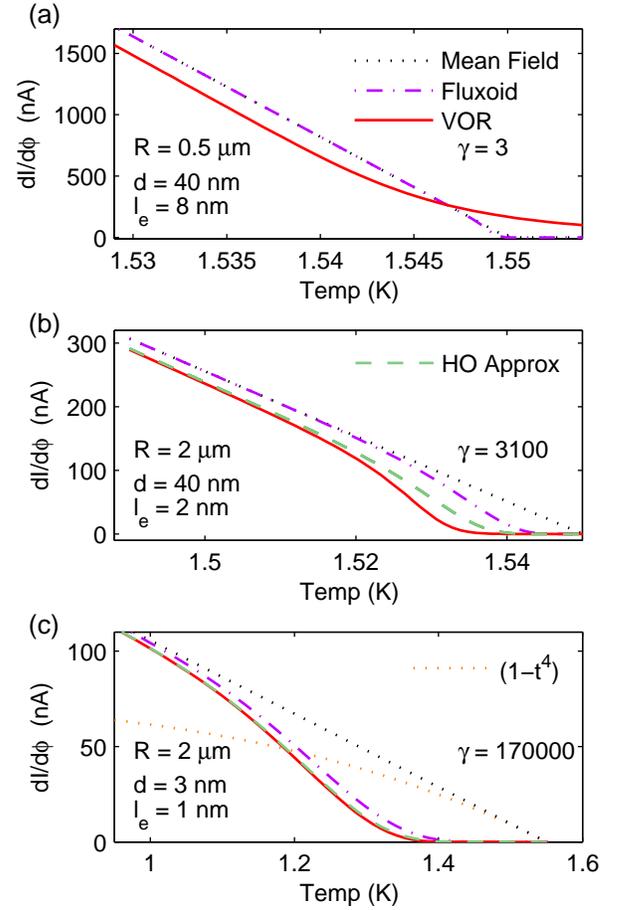}
\caption[Theoretical response from London, fluxoid, von Oppen, and HO models.]{ (Color Online) Theoretical susceptibilities calculated using the mean field
model, black dotted line Eq. (\ref{eq:Izero}), the fluxoid model,
purple dot-dash line Eq.
(\ref{eqn:suscfluxfluct}), the VOR model, red solid line
Eqs. (\ref{eq:thermocurrent}, \ref{eq:nVORpart},
\ref{eq:nVORhamiltonian}), and its approximate HO solution, green dashed
line Eqs. (\ref{eq:thermocurrent}, \ref{eqn:fullHO}), for rings with
$w = 80\,$nm and $T_c=1.55\,$K.
a) $\gamma(T=T_c) = 3$.  The VOR model predicts a susceptibility above $T_c$.
b) $\gamma(T=T_c) = 3100$.  A downturn occurs at
$dI/d\phi \approx 12 k_B T / \Phi_0 \approx 120 \rm{nA}$, $T\approx1.52,$K.  The
fluxoid, HO and VOR models reproduce the overall lineshape of
the downturn, up to an offset in $T_c$.  However, the three models
predict downturns of different sizes with the largest predicted by the
VOR model.
c) $\gamma(T=T_c) = 170,000$ Fluxoid fluctuations dominate the response over a wide
temperature range and the HO and fluxoid models
become increasing accurate predictors of the full fluctuation theory.
For all values of $\gamma(T=T_c)$,
the VOR and HO response well below $T_c$ only match
the mean field and fluxoid predictions if $T_c$ is renormalized.
\label{fig:von_ho_fluxf}}
\end{figure}

\subsection{Harmonic Oscillator Model}
\label{sec_HO}
The harmonic oscillator (HO) approximation is valid at temperatures
well below $T_c$, where the wave functions contributing
to Eq. (\ref{eq:nVORpart}) only extend over a narrow region around
the minimum of the Mexican hat potential of Eq. (\ref{eq:nVORhamiltonian}),
so that the latter can be approximated by a quadratic expansion.
In this case, fluctuations from the quartic nature of the
potential should not play a significant role.
We refer to the fluctuations in this model as quadratic fluctuations,
rather than Gaussian fluctuations,
to avoid confusion with small order parameter fluctuations above $T_c$,
which are often referred to as Gaussian fluctuations.

Eigenstates have the
form ${\vec r} = |r|\exp(i l \phi)$, so Eq.
(\ref{eq:nVORhamiltonian}) can be written as a 1D problem,
$H = - \frac{1}{2} \frac{d^2}{dr^2} + V(r)$ where
\begin{equation}
V(r) = \frac{l^2}{2 r^2} + \frac{1}{2}\frac{\Lambda(T)^2}{\gamma(T)^{2/3}} {r}^2 + \frac{1}{4} {r}^4.
\label{eqn:voppVff}
\end{equation}
Expanding $V(r)$ about its minimum at $R_{\textrm{m}}(l)$ leads to the  eigenvalues
\begin{equation}
\mathscr{E}_{n,l}= \frac{l^2}{2R_{\textrm{m}}(l)^2} + \frac{R_{\textrm{m}}(l)^4}{4} + \omega (n + 1/2)
\label{eqn:HO_Enl}
\end{equation}
where $\omega = \sqrt{\Lambda(T)^2/\gamma(T)^{2/3} + 3 R_{\textrm{m}}(l)^2 + 3 l^2/R_{\textrm{m}}(l)^4} $.

Only terms that change with $l$, the angular momentum coordinate in
the fictitious Hamiltonian, contribute to the flux dependence of
the partition function, thus only these terms contribute to the
thermodynamic ring-current.
If we make an approximation and only include the $l^2/2R_{\textrm{m}}(0)^2$ terms,
where $R_m(0)$ is the value for $r$ that minimizes $V(r)$ when $l=0$,
the current from
Eq. (\ref{eq:thermocurrent}) is
\begin{eqnarray}
\lefteqn{I_{HO}(T,\phi) = }\nonumber\\
& & {\frac{k_B T}{\Phi_0} \frac{\displaystyle\sum_{l=1}^{\infty}4\pi{l}\sin{(2\pi{l}\phi)}\exp(l^2\gamma(T)/2\Lambda(T)^2)}{1+\displaystyle\sum_{l=1}^{\infty}2\cos{(2\pi{l}\phi)}\exp(l^2\gamma(T)/2\Lambda(T)^2)}},
\label{eqn:simpleHO}
\end{eqnarray}
which is exactly equivalent the fluxoid current shown in Eq.
(\ref{eqn:fluxfluct}). Through this approximation we are able
to show a direct link between the harmonic oscillator approximation
to the VOR model and the fluxoid model.
Including the second two terms of Eq. (\ref{eqn:HO_Enl}), which account for the angular momentum dependence of $\omega$ and $R_{\textrm{m}}(l)$, we get
{\setlength\arraycolsep{2pt}
\begin{eqnarray}
Z & = & \sum_{l = -\infty}^{\infty} \exp(-i 2 \pi l \phi)  \exp(- \gamma(T)^{1/3} V(R_{\textrm{m}}(l))) \nonumber\\
    && {} \times \frac{\exp(- \gamma(T)^{1/3} \omega/2)}{1- \exp(- \gamma(T)^{1/3} \omega)}.
\label{eqn:fullHO}
\end{eqnarray}}

Using this simplified partition function we can find the ring's current
and consequently its susceptibility in the limit where we ignore only
quartic fluctuations.

\subsection{Comparison of Models}
\label{sec_comp}
We have presented the theoretical basis for four models
including: the mean
field model, the fluxoid model, the harmonic oscillator model and the
von Oppen and Riedel model. We now compare the physics captured by
each model by plotting the theoretical susceptibility
response predicted by each model as a
function of temperature for rings with
three different $\gamma(T=T_c)$
parameters in Fig. \ref{fig:von_ho_fluxf}.

The mean field model is our baseline.  It gives the ring response
in the absence of all superconducting fluctuations.  At the
other extreme, the VOR model incorporates all thermally activated
superconducting fluctuations into its
derivation of the ring response.
In between we have the fluxoid model, which includes only
fluxoid fluctuations and the harmonic oscillator model.
By comparing
these models for rings with different $\gamma(T=T_c)$ we can get a
sense of which fluctuations dominate the response.

One striking feature in Fig. \ref{fig:von_ho_fluxf}
in all three plots for all values of $\gamma(T=T_c)$,
is that both the VOR model and its HO
approximation have an offset in the linear regime, far below $T_c$,
compared to the mean field or fluxoid model.
This
downshift in $T_c$ appears to reflect a
renormalization in $T_c$ due to consideration of all possible
fluctuation modes.

Fig. \ref{fig:von_ho_fluxf}(a) shows a ring with $\gamma(T=T_c) = 3$.
The low gamma parameter means it has a strong phase
stiffness. 
The susceptibility at $T = T_c$ is 210 nA.  We get this number
by looking at the value of the
VOR model at $T_c$ in Fig. \ref{fig:von_ho_fluxf}(a).
We compare this number to $12k_BT/\Phi_0\approx120\,$nA,
for $T=1.5\,$K  predicted as the point
where we expect a downturn in susceptibility due to fluxoid fluctuations,
Eq. (\ref{eq:ff_limit}).  Since the fluxoid criterion is
smaller that the susceptibility at $T_c$ , susceptibility enhancing
amplitude fluctuations at and above $T_c$
overwhelm the susceptibility reduction expected from fluxoid fluctuations.
A downturn is not observable, instead the small $\gamma(T=T_c)$
leads to a susceptibility signal above $T_c$.

When $\gamma(T=Tc) = 3100$, as shown in Fig. \ref{fig:von_ho_fluxf}(b),
the fluxoid induced downturn becomes visible below $T_c$
starting at $T\approx1.52\,$K and 120 nA,
as predicted by our fluxoid criterion, Eq.
(\ref{eq:ff_limit}).  All three fluctuation models qualitatively
reproduce the shape of the susceptibility suppression.
As expected, VOR model predicts
a greater susceptibility suppression than the fluxoid
or HO models, because only the VOR model includes all thermal fluctuations.
The excess suppression between the fluxoid and VOR
is presumably due to contributions
from non-homogeneous phase winding solutions, amplitude fluctuations, or both.
While the excess suppression between the HO and VOR models is due
to fluctuations caused by the quartic nature of the potential.

For $\gamma(T=T_c) = 170,000$, shown in Fig. \ref{fig:von_ho_fluxf}(c), the
susceptibility response is dominated by fluxoid fluctuations,
shown by the almost identical lineshape shared by the fluxoid model
and the VOR model. The total response is also well represented
by the harmonic oscillator approximation showing that in
this region nearly all fluctuations are quadratic in nature.

Fig. \ref{fig:von_ho_fluxf}(c) shows a
larger temperature range than the previous panels,
and the GL approximation that $T$ is close
to $T_c$ is not valid over the whole plot.
GL theory is strictly valid in
the range of temperature where the
linear mean field response approximates
a temperature dependence that goes as $(1-t^4)$, $t=T/T_c$, shown
as an orange dotted line.  An alternative criterion is that $T>\Delta(T)$,
where $\Delta(T)$ is the superconducting gap.  These both result in
approximately the same range of validity.
GL theory has been applied with success at temperatures far
from $T_c$, but interpretation of results in this regime should
be treated with caution. The $(1-t^4)$ dependence is not included in panels
(a) and (b) because all plotted temperatures lie within the valid range.

In the next section we describe our measurement of ring susceptibility
for rings with different $\gamma(T=T_c)$ values.  We find good agreement
between our data and the fluctuation response predicted
by the fluxoid and VOR models.

\section{Sample and Measurement Technique}
\label{sec_meas}

\subsection{Sample Preparation}
We measured quasi-one-dimensional superconducting rings in a dilution
refrigerator \cite{BjornssonPG:Scasqi} with a scanning SQUID susceptometer
\cite{huber_gradiometric_2008} that was explicitly designed for this purpose.
We focus on data from two different samples expected to exhibit fluxoid
fluctuations.  Sample I's rings were fabricated and measured previously.\cite{bluhm_magnetic_2006}
The rings were narrow and
dirty with ${T_{c}}_{\rm{I}}\approx 1.5$ K.  They were made by depositing
a 40 nm thick Aluminum film by e-beam evaporation at a
rate of about $1\,\rm{\AA}$/s and a pressure of approximately
$10^{-6}$ mBar on a Si substrate patterned with poly(methyl methacrylate) (PMMA) resist. During the deposition, the rate temporarily
dropped to a negligible level for about 10 min and subsequently
recovered. This delay caused the formation
of two superconducting layers separated by an AlOx
tunneling barrier.  The coupling between the two Al layers depended on
the width of the rings with narrow rings ($w\leq190\,$nm) and wide rings
($w\geq250\,$nm) showing a single order parameter.  Intermediate widths
showed evidence of weak interactions between the two layers leading to
two order parameter effects.\cite{bluhm_magnetic_2006}
In this work we only present data from the
narrow rings which showed no two order parameter behavior.
However, due
to the oxidization process we suspect the thinnest rings have a large
oxidized layer that reduces the thickness of the superconductor.
Consequently we expect that these rings have an effective height
that is less than $40\,$nm.  We can test this prediction by extracting the
ring's cross-section from fits to the VOR model.

The rings on sample II were fabricated specifically for this paper.
The fabrication process was almost identical to the rings
from sample I except the evaporated film was thinner,
$d=15\,$nm, and there was no interruption in the evaporation.
The deposited rings were
wide and dirty with ${T_{c}}_{\rm{II}}\approx2.1\,$K. Of the
many fabricated rings of different widths and radii, only the widest rings,
$w\approx850\,$nm, had a diamagnetic
response.  The next widest rings, $w\approx 450\,$nm, showed no signs of
superconductivity indicating that they were oxidized throughout.  This
evidence makes it difficult to predict with certainty what portion, if any,
of the $850\,$nm rings are also oxidized.  Although the 1D
approximation, $w>>\xi(T)$, is no longer strictly valid for these rings
the theory still provides a good fit to our data.
For each sample we used
Ginzburg Landau models\cite{tinkham_SC,Zhang,bluhm_magnetic_2006}
to fit a zero temperature penetration depth
$\lambda_{\rm{I}}(0) \approx 800$ nm, $\lambda_{\rm{II}}(0) \approx 1.5$ $\mu$m,
and coherence length $\xi_{\rm{I}}(0) \approx 80$ nm,
$\xi_{\rm{II}}(0) \approx 30$ nm.

\subsection{Measurement}
Our measurements are done with a voltage biased DC SQUID
susceptometer amplified by a series-array
SQUID preamplifier.\cite{SAS}  The SQUID is mounted on a piezo-resistive
scanning assembly\cite{BjornssonPG:Scasqi} which is connected to
the mixing chamber of a dilution
refrigerator through a single copper braid.  The temperature of the scanner
and sample is controlled with sub-milikelvin precision though feedback.
The SQUID sensor's counter-wound geometry, with on-sensor modulation
coils for feedback, enable cancellation of an applied field to one part in
$10^4$.\cite{huber_gradiometric_2008}
The ring current is measured by positioning the SQUID about $1\,\mu$m
above the ring and recording the flux induced by the ring's current
in the SQUID's $4.6\,\mu$m diameter pick-up loop.
During the measurement, the applied flux threading the ring is varied by
several flux quanta at a few Hertz by an on-sensor field coil.
This measurement is repeated $13\,\mu$m
above the ring and the ring signal is computed as the difference between the
two positions for each value of applied flux.  This procedure allows
us to achieve an additional three orders of magnitude of background
cancellation.  A more detailed description
of the measurement system was given by
Koshnick \textit{et al}.\cite{koshnick_fluctuation_2007}

\begin{figure}[t]
\centering
\includegraphics[scale=1]{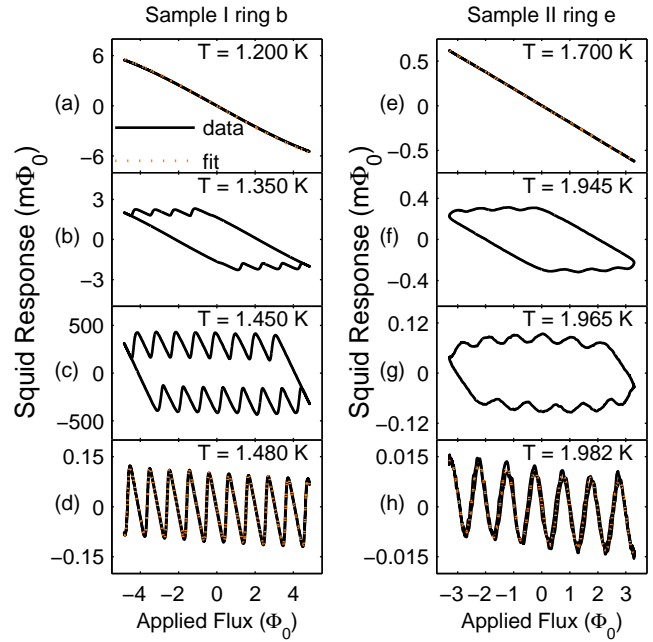}
\caption[$\Phi-I$ Curves ]{
(Color Online) Plots of SQUID response vs. applied flux at different
temperatures for a ring from each of the two samples.
Rings b and e refer to specific
rings plotted in Fig. \ref{fig:suscdata}. Ring b $T_c = 1.56\,$K and
ring e $T_c=2.08\,$K.  The curves evolve from
non-hysteretic with no fluxoids at low temperatures, through a
hysteretic regime, to non-hysteretic with a change in fluxoid number
at every applied flux quantum near $T_c$.
The orange dotted lines are fits to the GL current,
Eq. (\ref{eq:GL_current}) at
low temperatures and to the Boltzmann distribution,
Eq. (\ref{eqn:fluxfluct}), at high temperatures.
We extract the ring's susceptibility at each temperature by taking the
derivative at $\phi=0$.
\label{fig:phiIcurves}}
\end{figure}

We plot the flux induced in the SQUID's pick-up loop as a function
of the flux applied by the field coil in Fig. \ref{fig:phiIcurves}
for two different rings. The measurement is repeated
to record the full temperature dependence of the ring's response.
The ring current, $I$, is coupled as flux into the SQUID pick-up loop
through the mutual inductance, $M$.
$\Phi_{\rm{SQUID}}=MI$.  We estimate the mutual inductance between the
SQUID pick-up loop and
a ring by calculating the mutual inductance between two on axis rings with
radii $r_1$ and $r_2$ a distance $z$ apart.
\begin{equation}
M = \pi\mu_0{r_1}{r_2}\int_{0}^{\infty} d\kappa e^{-\kappa|z|} J_1(\kappa{r_1})J_1(\kappa{r_2})
\label{eq:MI}
\end{equation}
$J_1$ is a Bessel function of the first kind.
For all our mutual inductance calculations
we assume a ring-pick-up loop separation of $1\,\mu$m.
Through a separate fitting technique\cite{bluhm_magnetic_2006}
we estimate the actual distance between the pick-up loop and
the ring to range from $0.75-1.1\,\mu$m. Ring currents and
susceptibilities quoted later in this paper have error bars that
reflect this systematic uncertainty in the coupling factor which
would shift the entire dataset.
For example, the ring-SQUID mutual inductance between a ring
with a radius of $1.5\,\mu$m positioned $1\,\mu$m above the
SQUID's pick-up loop is $M=0.718\,\Phi_0/$nA.
The error on the height from $0.75-1.1\,\mu$m
gives $M = 0.669-0.849\,\Phi_0/$nA, but because the SQUID is
kept a constant distance from the ring  for all
data points taken at the same time on the same ring this systematic
error in the height would shift all data points together.

The ring response curves plotted in Fig. \ref{fig:phiIcurves} evolve from
cubic and non-hysteretic at low temperatures through a hysteretic regime to
periodic and non-hysteretic near $T_c$.  At low temperatures the current
response is well described by the GL current with no phase
windings, Eq. (\ref{eq:GL_current}) with $n=0$,
shown as a orange dotted line in panels (a) and (e) of
Fig. \ref{fig:phiIcurves}.
As the temperature increases the
applied flux causes the ring
to transition to higher fluxoid states;
however, the phase slip rate becomes comparable
to the measurement time only at large applied flux, thus leading
to a hysteretic response.

Finally, as the temperature approaches $T_c$ the phase slip rate becomes
fast compared to the measurement time at small applied flux and the
ring relaxes to thermal equilibrium.  Thermal
fluctuations are strong enough for
transitions to occur within some range of $\phi = m/2$
where $m$ is an odd integer.  The ring's response is no longer
hysteretic and can be approximately modeled as a Boltzmann
distribution of all fluxoid  states, Eq. (\ref{eqn:fluxfluct}),
shown as a orange dotted line
in panels (d) and (h) of Fig. \ref{fig:phiIcurves}.
We extract the magnetic susceptibility of the ring at each temperature by
fitting a low order polynomial to obtain the slope at $\phi=0$.

\subsection{Susceptibility Data}
\label{sec_susc}
\begin{figure}[h!]
\centering
\includegraphics[scale=.99]{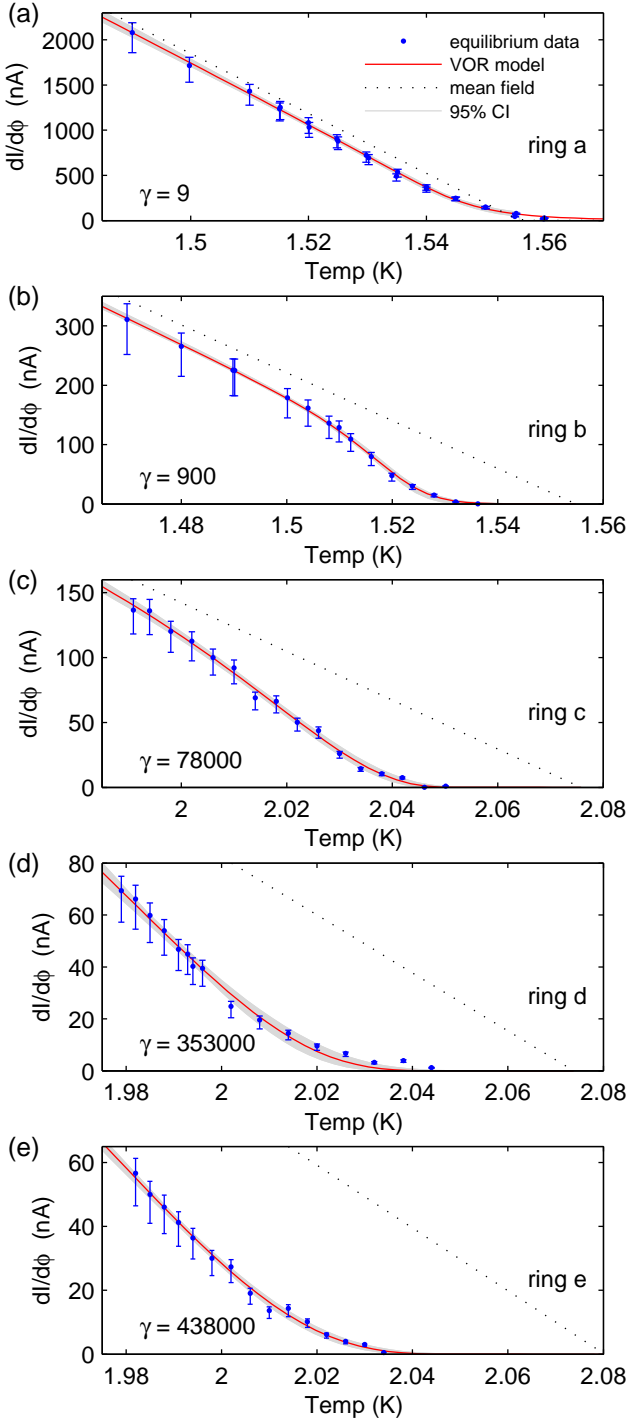}
\caption[Data of fluxoid fluctuation reduced response.]{
(Color Online)
Zero field susceptibility (blue dots) vs. temperature plotted
for five different rings. In all plots the solid red line is a
fit to the VOR model, Eqs. (\ref{eq:nVORpart}, \ref{eq:nVORhamiltonian}).
The dotted black line is the expected mean field susceptibility
given by $I^0(T)$, Eq. (\ref{eq:Izero}).  The gray shaded area
represents the 95\% confidence interval from bootstrapping.
The error bars represent the systematic uncertainty in the
SQUID-ring mutual inductance.
\label{fig:suscdata}}
\end{figure}

We measured thirty-eight rings on sample I and twelve rings
on sample II.  Sample I was fabricated and measured primarily for a
different experiment.\cite{bluhm_magnetic_2006}
As a result only eight of the rings measured
have sufficient susceptibility data over a wide enough temperature range
to test the theories presented in the previous section.  Two
representative rings were selected for this paper.  The three rings
from Sample II were chosen to show a variety of ring parameters, and
because they had the most dense susceptibility data over the important
temperature range.  The set of five rings
allows us to explore the effects of ring size and
cleanliness on the fluctuation response. Fig. \ref{fig:suscdata} shows
the susceptibility vs. temperature data for those five rings. Each of the ring's
physical parameters are given in table \ref{tab:data}.
We extracted the ring radii from the flux periodicity of the ring's
response in thermal equilibrium and confirmed the measurement
though SEM imaging.
The ring thicknesses were measured with an AFM, and the width with SEM.
Fitting to the VOR model allowed us to estimate values
for the ring's cross-section and mean free path.
We used the measured ring width and thickness plus an additional error factor
as an upper limit on
the cross-section parameter in the VOR model for rings (c-e).
No lower limit was enforced
due to the possibility of oxidation reducing the superconducting cross-section.
\newcommand{\rb}[1]{\raisebox{1.5ex}[0pt]{#1}}
\begin{table}[h]
\begin{tabular}{|l|r|r|r||r|r|r|r|r|}
\hline
 & \multicolumn{3}{|c||}{Directly Measured} & \multicolumn{4}{|c|}{Extracted from VOR fits} \\
\hline
\multirow{2}{*}{Ring} & $R$ & $w$ & $d$ & $wd$ &$95\%$ CI & $ \ell_e$& $95\%$ CI\\
 & ($\mu$m) & (nm) & (nm) & (nm$^2$) &  & (nm) & \\
\hline\hline
a(I) & 0.496 & 123 & 40 & 1598 & 1140-2314 & 6.4 & 4.3-9.4\\
b(I) & 1.97 & 90 & 40 & 583 &492-1177 & 8.5 & 4.4-10.4 \\
c(II) & 1.21 & 840 & 15 & 13319 & 2811 -14790 &  0.11 & 0.09-0.48\\
d(II) & 1.75 & 850 & 15 & 14790 & 11896-14790 &  0.08 & 0.07-0.11 \\
e(II) & 1.82 & 850 & 15 & 13602 & 9172 -14790 &  0.08 & 0.07-0.12 \\
\hline
\end{tabular}
\caption{Table of ring values.  Values for the cross-section
and mean free path, extracted from fits to the VOR model,
are given with their $95\%$ confidence interval. An upper limit of
$14790\,$nm$^2$ was enforced on the ring cross-section
to constrain the fits for rings (c-e).
\label{tab:data}}
\end{table}

\begin{table}[h]
\begin{flushleft}
\begin{tabular}{|l||r|r|}
\hline
Ring & $T_c$(K) & $95\%$ CI \\
\hline
\hline
a(I) & 1.556 & 1.554-1.557 \\
b(I) &  1.555 & 1.550-1.556\\
c(II) & 2.076 & 2.072-2.086\\
d(II) &  2.074 & 2.066-2.083\\
e(II) &  2.080 & 2.075-2.086\\
\hline
\end{tabular}
\begin{tabular}{|l||r|r||r|r|}
\hline
Ring & $\gamma(T=T_c)$ & $95\%$ CI & $E_0(T=0)$ &$95\%$ CI \\
\hline\hline
a(I) & 9.20  & 6-14 &  {2138} & 1850-2550 \\
b(I) & 899 & 760-1660 & {898} & 820-1300 \\
c(II) & $78\times{10^{3}}$ & $(18-95)\times{10^3}$ & 2650 & 1180-2870 \\
d(II) & $353\times{10^3}$ & $(261-443)\times{10^3}$ & 2590 & 2280-2730 \\
e(II) & $438\times{10^3}$ & $(294-536)\times{10^3}$ & 2370 & 1910-2550\\
\hline
\end{tabular}
\caption{Table of fitted values. We used the temperature independent
portions of $\gamma(T)$ and $E_0(T)$ as fit parameters
in the VOR model.
This table reports values for $\gamma(T=T_c)$ and $E_0(T=0)$
as well as the limits of the $95\%$ confidence interval
obtained from bootstrap analysis for
the data presented in Fig. \ref{fig:suscdata}.
\label{tab:results}}
\end{flushleft}
\end{table}

Fig. \ref{fig:suscdata} plots the susceptibility vs. temperature
curves for five rings. The blue susceptibility data
points represent the slope at $\phi=0$ of the SQUID response at
different temperatures scaled by
the ring-SQUID mutual inductance to get the ring current.  The error
bars represent height errors in our calculation of the mutual inductance,
Eq. (\ref{eq:MI}). This error is systematic and expected to be the same
for all points in a panel.
Using $T_c$, and the temperature independent portions of
$\gamma(T)$ and $E_0(T)$ as the free parameters, the red line is a fit
of the data to the VOR model, Eqs.
(\ref{eq:nVORpart}, \ref{eq:nVORhamiltonian}).  The fit results used to
generate the red curves are given in table \ref{tab:results}.
We report values for $\gamma(T)$ at $T_c$ and $E_0(T)$ at $T=0$.
The reported $T_c$ represents the nominal mean field $T_c$
entering the VOR model.\cite{vonOppen}
The fitted values of $\gamma(T=T_c)$ are also listed on each of the plots.
The black dotted line is the mean field
ring response, Eq. (\ref{eq:Izero}),
which one would expect if no fluctuations were present.  Deviations
in the data from the black dashed line show the
influence of fluctuations on a given ring.  Finally, the gray
region of the curve is the $95\%$ confidence interval (CI)
obtained from bootstrapping.

Using the fit results from table \ref{tab:results} along with
the known values
of the ring radii given in table \ref{tab:data} we can extract
values for the ring's cross-section and mean free path from
expressions for $E_0(T)$,
Eq. (\ref{eq:E0}), and $\gamma(T)$, Eq. (\ref{eq:gamma}).
The ring parameters obtained
in this way are give along with their
$95\%$ confidence intervals in table \ref{tab:data}.
Due to the evaporation conditions discussed
previously, we're not confident that the entire cross-section
of each ring is superconducting. For the two rings on sample I
the fitted cross-sections are
smaller than the values found using AFM/SEM, which confirms our
suspicion that a portion of the ring is oxidized.
The data from the three
sample II rings is within the downturn region, ie the
decrease in the susceptibility is not linear even at the lowest
plotted temperatures.
Practically we are limited on the low end of the
temperature range by the point
where the SQUID response curves go hysteretic.
A three parameter fit is under-constrained and it is
consequently difficult to get accurate VOR fits
without susceptibility
data at lower temperatures including the point where the
data is reduced from the linear response.

As a result,
for rings (c-e) we put a strict upper
limit of $w\times{d}=14790\,$nm on the cross-section, which
acted as an additional constraint on the VOR fits.
The cross-sections extracted from fits to the constrained VOR model
for the rings on sample II agree well with the
AFM/SEM cross-section indicating little oxidation.
A similar limit was not
applied to rings (a-b) because data in the linear susceptibility
region kept the fit from being under-constrained.

The agreement between the
susceptibility data and fits to the VOR model
are good for all rings except ring (d), where  it is clear
that the VOR model does not capture the shape of the data at
high temperatures.
It is unclear why the VOR model provides a poor fit for this ring.
It is possible that errors from extracting the susceptibility
near $T_c$, errors
that are not accounted for in the error bars, are particularly
large for measurements on this ring.

Looking at the sequence of five rings
it is clear that the extent of the suppression
of superconductivity increases as $\gamma(T=T_c)$ increases.  This is
just what we expect for a series of
rings where fluxoid fluctuations play a larger and larger role.

Ring (a) shows an enhancement
of the superconducting response above $T_c$.
This response is caused by amplitude fluctuations and has been studied by
Koshnick \textit{et al.} \cite{koshnick_fluctuation_2007} and Zhang and
Price.\cite{Zhang}  As we showed in our description of the theoretical
models only the VOR model can correctly
reproduce the upturn in susceptibility above $T_c$.
The fluxoid fluctuation model cannot reproduce a current above $T_c$ and 
will fail for any ring with a $\gamma(T_c)<40$.

The remaining four rings in Fig. \ref{fig:suscdata} show a suppression of
the susceptibility signal below the mean field response (shown in black dots).
However, of the plotted rings
only ring (b) has a large enough temperature range to observe
a downturn from the linear regime.  The full temperature range plotted for
rings (c-e) is already deep in the suppression region.  
This is due to the fact that $\gamma(T_c)>16,000$ for rings (c-e).
In the next
section we expand the temperature range by adding susceptibility data
from lower temperature hysteretic ring response curves, to confirm
that the response is suppressed from the mean field value.
The downturn for ring (b) occurs at $T\approx1.52\,$K and
$120\,$nA, which corresponds to the
criterion for fluxoid fluctuations given in Eq. (\ref{eq:ff_limit}).
Such agreement validates our criterion for the onset of
susceptibility suppression driven by fluxoid fluctuations.

\begin{figure}[t]
\centering
\includegraphics[scale=1]{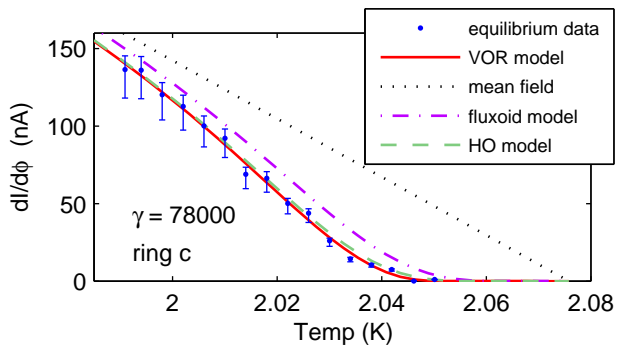}
\caption[Data of fluxoid fluctuation reduced response.]{
(Color Online)
Comparison of the four models plotted using parameters obtained from
fitting the data from ring (c) to the VOR model.
Fitting to the fluxoid model would yield an equally good fit with
a slightly different $T_c$.
\label{fig:112}}
\end{figure}

We have shown that the VOR model describes the temperature
dependence of the susceptibility.  To get a feeling for
the type of fluctuations that play a role in the ring response
we plot the fluxoid model and the HO model in addition to the
VOR model and mean field model for ring (c) in Fig. \ref{fig:112}.
It is clear that fluxoid fluctuations cause the majority of the
suppression, with quadratic fluctuations described by the HO model
contributing to the renormalization of $T_c$
and quartic fluctuations described by the VOR model
playing only a minor role.  In fact the fluxoid model
would fit the data equally well with just a shift
in the $T_c$.

The dataset, taken as a whole, confirms the points we made
throughout this paper.  Fluxoid fluctuations not only
suppress the rings superconducting response but play an
increasingly large role in the suppression
as $\gamma(T=T_c)$ increases.  We showed that our susceptibility
vs. temperature data is well described by a GL model
for homogeneous rings, developed by
von Oppen and Riedel,\cite{vonOppen}
that includes all thermally activated fluctuations, but
the largest gamma rings can be equally well described by
our simple fluxoid only model with a shifted $T_c$.  Furthermore
we can use fits to the VOR model to reproduce some of the rings
physical parameters including the cross-section and mean free path.
Finally, by using VOR fit parameters we can
employ our two approximate models, the fluxoid model, and the
harmonic oscillator model, to determine the how much of the suppression is
due to either fluxoid fluctuations or quartic fluctuations,
done for ring (c) in Fig. \ref{fig:112}.

\subsection{Hysteretic Susceptibility Data}
\label{sec_hyst}
For rings (c-e) in Fig. \ref{fig:suscdata} we expect the onset of the
downturn induced by fluxoid fluctuations in a
temperature range where the SQUID response curves
are hysteretic, as shown in Fig. \ref{fig:phiIcurves}.
This is due to the fact that in these thinnest,
dirtiest rings $L>>\xi(T)$ and fluxoid fluctuations
are already energetically
favorable at the temperature when phase slips begin to occur,
as discussed in Sec. \ref{sec_phaseslips}.
Fluxoid fluctuations are never energetically favorable
for ring (a) and they onset well after phase slips
in ring (b). Phase slips onset at $\sim1.3\,$K while
fluxoid fluctuations onset at $\sim1.51\,$K.

\begin{figure}[h!]
\centering
\includegraphics[scale=1]{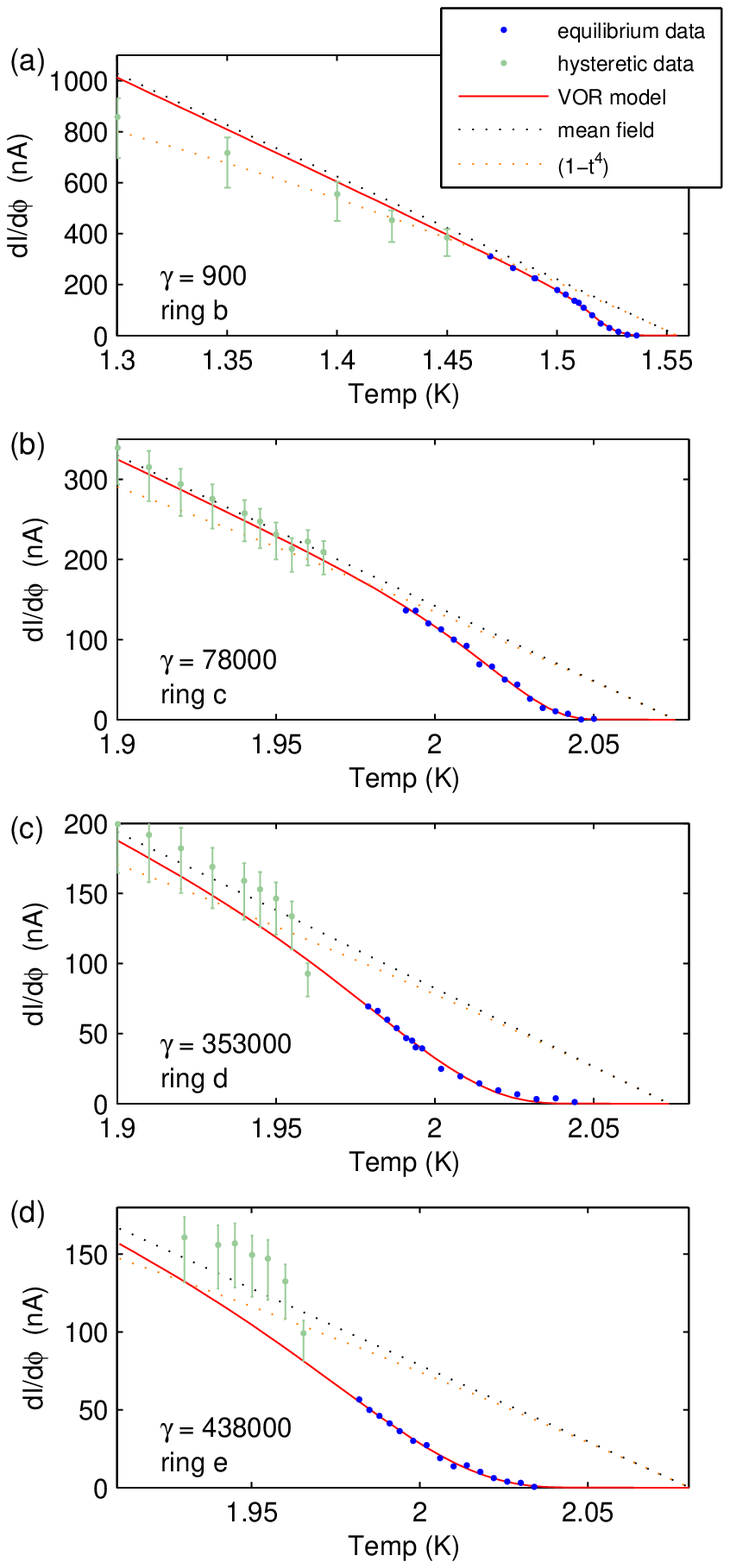}
\caption[Hysteretic Data ]{
(Color Online) Susceptibility data from rings (b), (c), (d) and (e).
The green points represent the slopes of the hysteretic
curves, which estimate the susceptibility in the hysteretic regime.
The error bars account for a systematic error in the coupling constant
that would shift all points together.
The blue points give the susceptibility of
the ring's response in thermal equilibrium.
The red solid line is a fit of the blue non-hysteretic 
data to the VOR model and the
black dotted line is the mean field response.  Also plotted is the
$(1-t^4)$ temperature dependence which places approximate
limits on the validity of GL.
\label{fig:hystdata}}
\end{figure}

To demonstrate that the data presented represents
a real reduction in the ring response we examine
the susceptibility signals at
lower temperatures that fall in the hysteretic regime.
We evaluate susceptibility in the hysteretic regime by taking the slope
at zero current on the long continuous sides of the hysteretic curves.
These susceptibility data points are shown as
green dots in Fig. \ref{fig:hystdata}.

Fig. \ref{fig:hystdata} shows susceptibility data in the hysteretic regime
(green points)
and reproduces the susceptibility data from the non-hysteretic regime
(blue points) from rings (b-e) in Fig. \ref{fig:suscdata}.
Fig. \ref{fig:hystdata} also shows the $(1-t^4)$ dependence,
plotted as an orange
dotted line, that sets the validity of our GL based models.
In Panel (a) the susceptibility in the hysteretic regime has dropped
below the mean field prediction and instead closely follows the
$(1-t^4)$ curve.

Notice that the hysteretic data points in rings (c-e), panels (b-d),
follow the mean
field curve until a crossover point when they line up with the VOR model
and the higher temperature susceptibility data.
This provides evidence that the
susceptibility measured from the
SQUID response curves in thermal equilibrium is suppressed
from the mean field value.  The drop in susceptibility from
the mean field value occurs when
phase slips occur at a sufficiently high rate and multiple fluxoid
states compete to suppress the response.

\section{Conclusions}

Superconducting phase slips in one dimensional rings and wires have been the
subject of theoretical and experimental interest for decades.
While phase slips in 1D structures determine the onset of resistance,
the fluxoid processes we
described here cause the loss of another hallmark of superconductivity, the
ability to screen magnetic field.
In this paper we have outlined four models that describe the effects of
superconducting fluctuations on the susceptibility response in
rings.  We have shown that ring responses for rings with
various physical parameters can be characterized by
a model by von Oppen and Riedel for uniform rings
that includes all thermal fluctuations.
However, by comparing the models we can determine the types of
fluctuations that contribute to the response of a given ring.
We specifically found that for rings with weak phase stiffness
the ring response can be described using a fluxoid only model, indicating
that these types of fluctuations are the dominant cause of
suppression of the susceptibility signal.
One could imagine extending this ring
system to a weakly connected grid, linking our results to the
field of percolation superconductivity.
Additionally, achievable experimental conditions allow
fluxoid fluctuations to occur at temperatures down to 50 mK, which could
provide an experiment setup for examining the quantum mechanical behavior
of a 1D ring.\cite{matveev_persistent_2002}

\begin{acknowledgments}
This work was supported by NSF Grants DMR-0803974, DMR-0507931, PHY-0425897
and by the Packard Foundation. Work was performed in part at the Stanford
Nanofabrication Facility, which is supported by NSF Grant No. ECS-9731293,
its lab members, and industrial affiliates.  We express gratitude to Martin
Huber for assistance in SQUID design and fabrication.
\end{acknowledgments} 


%

\end{document}